\documentclass[sigconf]{acmart}

\pdfoutput=1
\usepackage{listings}
\usepackage{booktabs} 
\usepackage{tikz}
\usetikzlibrary{calc}
\usepackage{mdframed}
\usepackage{pgfplots}
\usepackage{pgfplotstable}
\usepackage{graphicx}
\usepackage{subcaption}
\usepackage{caption}
\usetikzlibrary{calc}
\usetikzlibrary{patterns}
\usetikzlibrary{shapes.multipart, arrows}
\usetikzlibrary{decorations.pathreplacing}
\usepackage{multirow}
\usepackage[PseudoCode]{algorithm}
\usepackage[noend]{algpseudocode}
\usepackage{color}
\usetikzlibrary{shapes,snakes}

\setcopyright{none} 
\acmYear{2018}

\def\codename{Kosto}
\def\inp{$\texttt{Input}$}
\def\output{$\texttt{Output}$}
\def\auxdata{$\texttt{AuxData}$}

\def\encinp{$\texttt{Enc}(k_{\mathcal{P}}, \ \inp)$}

\def\program{$\texttt{Program}$}

\def\progSM{$\texttt{Prog}_{SM}$}
\def\progKT{$\texttt{ProgKT}$}
\def\encprogKT{$\texttt{Enc}(k_{\mathcal{P}}, \ \texttt{ProgKT})$}
\def\BAM{${\texttt{AM}}$}
\def\WKH{${\texttt{KH}}$}
\def\certB{$\texttt{Cert}_{\texttt{AM}}$}
\def\certW{$\texttt{Cert}_{\texttt{KH}}$}

\def\baseline{SimpleMarket}

\begin{document}\sloppy

\title{Fair Marketplace for Secure Outsourced Computations}

\author{Hung Dang}
\affiliation{
  \institution{National University of Singapore}
}
\email{hungdang@comp.nus.edu.sg}

\author{Dat Le Tien}
\affiliation{
  \institution{University of Oslo}
}
\email{dattl@ifi.uio.no}

\author{Ee-Chien Chang}
\affiliation{
  \institution{National University of Singapore}
}
\email{changec@comp.nus.edu.sg}

\maketitle

\section*{Abstract}

The cloud computing paradigm offers clients ubiquitous and on-demand access to a shared pool of computing resources, enabling the clients to provision scalable services with minimal management effort. Such a pool of resources, however, is typically owned and controlled by a single service provider, making it a single-point-of-failure. This paper presents \codename\ -- a framework that provisions a \textit{fair marketplace
for secure outsourced computations}, wherein the pool of computing resources aggregates resources offered by a large cohort of independent compute nodes. \codename\ protects the \textit{confidentiality} of clients' inputs as well as the \textit{integrity} of the outsourced computations and their results using trusted hardware's enclave execution, in particular Intel SGX. Furthermore, \codename\ warrants \textit{fair exchanges} between the clients' payments for the execution of an outsourced computations and the compute nodes' work in servicing the clients' requests. Empirical evaluation on the prototype implementation of \codename\ shows that performance overhead incurred by  enclave execution is as small as $3\%$ for computation-intensive operations, and $1.5\times$ for I/O-intensive operations.
\section{Introduction}
\label{sec:intro}
The cloud computing paradigm features a service provider offering its clients convenient and on-demand access to a shared pool of computing resources, enabling the latter to provision scalable services with minimal management effort. In such a model, clients typically entrust the cloud service provider to handle their data, to perform their outsourced computations, and to meter their cost footprint acurately (e.g., CPU-cycle, network bandwidth, storage)~\cite{cloud_accounting}. Although the clients may enjoy a wide range of cloud computing services, they are all offered by and at discretion of a few specific service providers.

Recent years have witnessed an emergence of online marketplaces that are in competition with traditional vendor-specific service providers. Examples include Airbnb~\cite{airbnb} in lodging, Uber~\cite{uber} in transportation, and Golem~\cite{golem} in outsourced computations. In such marketplaces, the shared pool of resources is no longer owned, provisioned and controlled by a single party, but rather aggregates those that are offered by a large cohort of independent providers. Designing a fair marketplace for secure outsourced computations, however, faces various challenges. 

The first challenge is in protecting the \textit{confidentiality} of the clients' data and the \textit{integrity} of the outsourced computations, for the resource providers (or compute nodes) may be untrustworthy. Solutions to protect the confidentiality and integrity of outsourced computations have been studied in the literature~\cite{goldreich_SMC, verifiable_computing, FHE09, FHE10}. For examples, homomorphic encryption~\cite{FHE09, FHE10} and secure multi-party computation~\cite{goldreich_SMC} are designed to protect data confidentiality, while verification by replications~\cite{golem, setihome} and verifiable computation~\cite{verifiable_computing} aim to protect computation integrity. Nevertheless, these approaches either incur prohibitive overheads or support only a limited range of applications, hindering their adoption in practical systems.

Another challenge is in codifying \textit{fair exchanges} between clients' payments for the execution of the outsourced computations and compute nodes' work in servicing the clients' requests. In the ``pay-as-you-use'' metering model where clients are billed based on the computing resources that their requested services consume, both clients and compute nodes have strong economic interests to falsify the resource metering (e.g., the compute nodes try to overcharge the clients, while the latter aim to be undercharged). One approach is to fix a remuneration for a task before hand, and let the compute node collect such reward once it returns a correct result for that task. To guarantee fairness, the remuneration might be deposited into an escrow held between the client and the compute node, which automatically and anomously releases the reward to the compute node upon successful task completion, or returns it to the client after a certain time-out. This approach, however, does not generalize. In particular, with micro tasks that yield very small remunerations, the transaction fee (i.e., the cost to conduct the payment transaction, which is often unproportional to the transaction value) becomes an overhead. In case of macro tasks, compute nodes may inadvertently abort the tasks midway (perhaps due to overly-extensive resource consumption), spending certain computational work but could not claim the reward. An ideal solution to enable fair exchanges between the clients and the compute nodes would require trusted metering of the latter's work and an self-enforcing or autonomous agent that is responsible for settling payments between the two parties based on the metering.

In this paper, we present \codename\ -- a framework that enables a fair marketplace for secure outsourced computations. Unlike in vendor-specific cloud services, a shared pool of computing resources in \codename\ aggregates a large cohort of independent \textit{compute nodes} each of which is capable of provisioning a Trusted Execution Environment (TEE) (e.g., using Intel SGX) for outsourced computations. The TEE provisioned by SGX, called an \textit{enclave}, prevents other applications, the operating system and even the host owner of the compute node from tampering with the execution of application loaded within or observing its state, thus guaranteeing data confidentiality and computation integrity. A client in \codename\ can request computational services from the compute nodes, while enjoying confidentiality protection on their data and integrity assurance on their outsourced computations. The compute nodes service the clients' requests by executing the outsourced computation inside an enclave that is attested to be correctly instantiated. Moreover, \codename\ enables fair exchange between the clients and compute nodes via a novel hybrid architecture that combines TEE-based metering with blockchain micro payment channel~\cite{raiden_eth, lightning_bitcoin}. More specifically, \codename\ incorporates in each enclave that houses the outsourced computation an accounting logic that correctly meters the compute node's work. Such metering is then translated to a {\em payment promise} with which the compute node can settle the escrow and claim the corresponding reward.
This allows the fair exchange between the client's payment and the compute node's work in executing the outsourced task, without incurring high overhead (i.e., transaction fee) or involving a trusted third party.

Besides, to facilitate the matching between clients and compute nodes, \codename\ designates brokers to collect resource advertisements from available compute nodes as well as task requests from clients. The brokers then evaluate among the requests and offers it receives appropriate assignments of tasks to compute nodes. We devise a solution for maximum task assignment in a dynamic settings wherein the broker continuously receives new requests from the clients and resource offers from the compute nodes. To eliminate monopoly of broker and avoid single-point-of-failure, \codename\ allows multiple brokers to co-exist, wherein a client or a compute node can also play a role of self-serving broker.

Empirical evaluations on the prototype implementation of \codename\ reveals that the overhead incurred by enclave execution and the trusted metering is reasonable, which is as small as $3\%$ for computation-intensive operations, and $1.5\times$ for I/O-intensive operations. 
We remark that while \codename\ is fully compatible with optimizations that enhance the efficiency of enclave execution~\cite{eleos, vault, hotcalls}, our prototype implementation does not include them. Thus, we expect the results reported in our evaluations (Section~\ref{sec:eval}) to be an \textit{over-estimation} of the real overhead that enclave execution incurs over unstrusted non-enclave execution, and the fully optimized implementation of \codename\ to attain better efficiency.

In summary, this paper makes the following contributions.
\begin{itemize}
\item We propose \codename\ -- a framework enabling a fair marketplace for secure outsourced computations that protects confidentiality of clients' inputs and integrity of the outsourced computations.

\item We codify in \codename\ a protocol that warrants fair exchanges between clients' payments for the execution of the outsourced computations and compute nodes' work in servicing the clients' requests.

\item We devise a task assignment mechanism that optimally matches pending requested tasks against available compute nodes in the system.

\item We conduct extensive experiments to demonstrate \codename's practicality. The experiments shows that performance overhead incurred by enclave execution and trusted metering is as small as $3\%$ for computation-intensive operations, and $1.5 \times$ for I/O-intensive operations.

\end{itemize}

The rest of the paper is organised as follows. We provide necessary backgrounds on Intel SGX, cryptocurrencies and their payment channels, as well as prior work on marketplace for outsourced computations in Section~\ref{sec:background}. Next, we give an overview of \codename\ in Section~\ref{sec:overview}, before presenting its design in Section~\ref{sec:design}. We then analyse its security in Section~\ref{sec:security_analysis} and conduct empirical evaluation in Section~\ref{sec:eval}. Finally, we survey the related works in Section~\ref{sec:related_works} before concluding our work in Section~\ref{sec:conclusion}.
\section{Preliminaries}
\label{sec:background}

\subsection{Trusted Execution Environment}
\label{subsec:sgx}

\codename\ leverages Intel SGX~\cite{sgx, sgx_remote_attest} to provision a Trusted Execution Environment (TEE) that protects confidentiality and integrity of the clients' data and computations. Nonetheless, we remark that \codename\ is compatible with any other mechanism featuring similar capabilities. In the following, we summarize key features of Intel SGX's TEEs.

Intel SGX~\cite{sgx} is a set of CPU extensions capable of providing hardware-protected TEE (or \textit{enclave}) for generic computations. It enables a host to instantiate multiple enclaves at the same time. An enclave is isolated from other enclaves, from the operating system (OS), and from the host itself. Each enclave is associated with a protected address space. The trusted processor blocks any non-enclave code's attempt to access the enclave memory. Memory pages can be swapped out of the enclave memory, but it is encrypted using the processor's key prior to leaving the enclave.

Intel SGX provides attestation mechanisms that allows an attesting enclave to demonstrate to a validator that it has been properly instantiated with the correct code~\cite{sgx_remote_attest}. In addition, the attestation mechanism also enables  the attesting enclave and the validator to establish a secure, authenticated channel via which they can securely communicate sensitive data. 

If the validator is another enclave instantiated on the same platform with the attesting enclave, the two parties engage in a {\em local attestation} mechanism. Abstractly, once the code in question has been initiated inside the attesting enclave, the trusted processor computes a \textit{measurement} of the initiated code (i.e., the hash of its initial state). Next, it produces a message authentication code (MAC) of such measurement using a key that is known only to the hardware and the validating enclave. Based on the measurement and the MAC, the validating enclave can verify if the attesting enclave has been instantiated correctly. Alternatively, if the validator is a remote party outside of the platform, the trusted processor creates a {\em remote attestation} by signing the attesting enclave's measurement with the hardware's private key under the Enhance Privacy ID (EPID) scheme~\cite{epid}~\cite{sgx_remote_attest}. The remote party obtaining the attestation then requests Intel's Attestation Service (IAS) to verify the signature contained in the attestation on its behalf~\cite{ias}. 

Recent attacks on Intel SGX show that enclave execution may be vulnerable to side-channel leakage~\cite{controlled_channel}, to which various defenses have been proposed~\cite{path_oram, oblivm, STC, tramer2017sealed}. \codename\ is compatible with these side-channel leakage defenses.

\subsection{Cryptocurrencies and Micro Payment Channel}
\label{subsec:payment_channel}

\noindent\textbf{Cryptocurrencies.} Decentralised cryptocurrencies have gained considerable adoption since the introduction of Bitcoin~\cite{btc_origin}. They are typically administered on the public ledger that is secured and maintained by a set of independent peer-to-peer network operators (or miners). Cryptocurrencies allow any two willing parties to transact directly with one another without relying on any trusted third party. While our discussion focuses on the Ethereum blockchain~\cite{eth_origin} due to its popularity and the expressive capability of its ecosystem, payment in \codename\ can be settled using any other currency that enables similar properties. 

\noindent\textbf{Smart Contract and Payment Escrow.}
Ethereum enables {\em smart contract} (or contract for short), which is an ``autonomous agent'' stored in the blockchain. A contract is associated with a predefined executable code. Incentive and security mechanisms of the Ethereum ecosystem encourage miners to execute the contract's code faithfully~\cite{eth_origin}. While various works have shown that miners could be incentivized to deviate from the contract code, they only apply to contracts that require nontrivial computation effort~\cite{luu2015demystifying}. 

Smart contract could be used to implement an \textit{escrow} that enables fair exchange between two parties without relying on a trusted third party. The escrow enforces a payment from a payer to a payee once the payee has delivered some service to the payer, while keeping the payment inaccessible to the payee before such delivery. \codename\ employs this capability to facilitate fair exchange between clients and compute nodes.

\noindent\textbf{Transaction Fee.} The contract implementing the escrow can be invoked by a transaction in the Ethereum network. Each such transaction is associated with a \textit{transaction fee} which is typically proportional to the execution complexity of the contract. In another words, the cost of settling the escrow does not necessarily depend on the monetary value that the escrow holds. Should the payment value is too small (i.e., micro transaction), the transaction fee becomes a significant overhead. 

\noindent\textbf{Blockchain Scalability.} The core component underlying cryptocurrencies is a consensus protocol which enables all (honest) peers in the network to ``agree'' on the same transaction history and to avoid double-spending. Consensus protocol in Ethereum's currently processes only a dozen of transactions per second\footnote{https://www.etherchain.org/}. Due to this limitations, \codename's design has to restrict the number of on-chain transactions, while still offering fine-grained resource metering and payment.

\noindent\textbf{Payment Channel.} Payment channel enables two parties to transact a large number of micro payments without incurring high transaction fee or overloading the blockchain with excessive number of transactions~\cite{raiden_eth, lightning_bitcoin}. A payer establishes a channel to a payee by depositing a maximum amount of value that they wish to be transacted into an escrow on the blockchain. Subsequently, when a payer wants to make a micro payment to the payee, instead of posting the transaction to the blockchain, he issues a digitally signed and hash-locked transfer, called \textit{payment promise}, and sends it off-chain to the payee. The value contained in the payment promises should not exceed the on-chain deposit that was set up previously, otherwise it cannot be fully collateralized. 

The escrow was programmed such that it can only be closed once using a single payment promise. The payment promises given to the payee have increasing value, where in each promise contains the sum of its immediate predecessor and the current micro payment. Let us assume the payer intends to make $n$ micro payments whose values are $\langle v_1, v_2, \cdots, v_n \rangle$. The $i^{th}$ payment promise takes the value of $\sum_{k=1}^i v_k$, invalidating all its predecessors. The payee can close the escrow at any time and claim the payment she has been promised so far by posting the last payment promise she received to the escrow. Upon receiving a closing payment promise, the escrow transfers the value indicated in the payment promise to the payee, and the remaining portion, if any, to the payer. 
This mechanism allows two parties to transact a very large number of micro payments using only two on-chain transactions (one for opening the channel, the other for closing it).

\subsection{Prior Work on Marketplace for Outsourced Computations: Golem}
\label{subsec:golem}
The most closely related project to \codename\ is Golem~\cite{golem}. The Golem network connects computers in a peer-to-peer network, and allows application owners and individual users to rent computing resources from other users' machines (i.e., compute nodes) to execute their tasks. Golem incorporates a dedicated Ethereum-based transaction system to facilitate direct payments between involved parties. Nonetheless, the current architecture and system designs of Golem face several challenges that are left unresolved.

First, Golem does not offer attested execution, allowing compute nodes to tamper with correct execution of outsourced tasks, or return bogus results. To verify the compute nodes' outputs, Golem has to repeat the same computation at different compute nodes, employing a majority voting principle to determine the correct result. This approach is not only subject to high overhead due to the redundant execution, but also susceptible to collusion wherein malicious compute nodes conspire to output the same incorrect result so as evade the voting.

Second, Golem does not feature trusted fine-grained metering. Compute nodes in Golem are only eligible to claim remuneration if they successfully finish the task and return correct results. If the assigned tasks are too large and they inadvertently have to abort midway (e.g., the task was ill-defined or consuming an unexpectedly large amount of resource), they cannot claim any reward despite having spent certain computation work. Further, a probabilistic nanopayment scheme in Golem\footnote{The payment scheme essentially runs the lottery among $n$ compute nodes $\langle {\mathcal C}_1, {\mathcal C}_2, \ldots, {\mathcal C}_n \rangle$, wherein ${\mathcal C}_i$ expects a payment of $v_i$. The lottery reward is $V = \sum_{i=1}^n v_i$, and the winning probability of participant ${\mathcal C}_i$ is $v_i/V$. The winning node collects the whole sum $V$, while others receive nothing.} requires compute nodes to actively involve in the system for a potentially long period of time before being able to collect its reward~\cite{golem_nanopayment}.

Finally, Golem requires the outsourced task to adhere to certain templates as defined in their task and transaction framework. This adherence is necessary for Golem's task verification and remuneration mechanisms.  Nonetheless, the expressiveness of Golem's task templates remains to be seen. 


\section{System Overview}
\label{sec:overview}
This section presents an overview of \codename, focusing on its architecture and desired properties. We then state the threat model and assumptions that we make in designing \codename. Finally, we introduce a strawman design that we shall convert step-by-step into \codename. 

\subsection{\codename\ overview}
\label{subsec:overview}

\codename\ is a framework that enables a fair marketplace for secure outsourced general-purposed  computations. Unlike vendor-specific cloud services, the pool of computing resources in \codename\ aggregates those offered by a large cohort of independent \textit{compute nodes} (which we shall discuss in more details below). Computation and data that are executed and processed in \codename\  enjoy strong confidentiality and integrity protections, thanks to the use of TEE on the compute nodes. Further, \codename\ ensures fair exchange between the clients and compute nodes via a novel hybrid architecture that combines TEE-based metering with the Ethereum blockchain\footnote{The \codename payment mechanism can also be settled using other currencies that feature similar properties of the Ethereum blockchain.}. The \codename\ architecture comprises three main parties: \textit{clients, compute nodes} and \textit{brokers}.

\begin{itemize}
\item \textbf{Clients} are end users of \codename. A client would like to the execute a program \program\ on an input \inp. The program \program\ can be written by the client, or an open-source software provided by a third party. In either case, the client outsources such computational task to a compute node. For practical usability reason, the clients are not expected to maintain constant connection with the compute node executing its program over the course of the outsourced computation.

\item \textbf{Compute nodes} are machines equipped with trusted processors (e.g., Intel SGX processors) that are capable of provisioning TEEs (or enclaves). A compute node services a client request by running its code in an enclave, and generating an attestation that proves the correctness of the code execution (and thus the result). In return, the compute node receives remuneration $v$ proportional to computational work it has asserted in executing the outsourced task.

\item \textbf{Brokers} facilitate node discovery as well as load balancing. Moreover, the brokers assist the clients in attesting that the compute nodes have correctly instantiated the enclaves housing the outsourced computations. Brokers may charge clients and/or compute nodes certain commission fee in return to their services. \codename\ eliminates broker monopoly and single-point-of-failure by allowing multiple brokers to co-exist, thus enabling better brokering service for both clients and compute nodes. 

\end{itemize}

\noindent\textbf{Work flow.} Hereafter, we denote by $\mathcal{P}$ a client, by $\mathcal{C}$ a compute node, and by $\mathcal{B}$ a broker. 
To achieve desired properties (discussed in Section~\ref{subsec:system_goals}), \codename\ requires \program\ to be instrumented into a program \progKT\ that incorporates dynamic runtime checks over the execution of \program. 
$\mathcal{P}$ and $\mathcal{C}$ can post their requests and available resource offers, respectively, to a broker $\mathcal{B}$ of their choice, perhaps based on the broker's reputation or quality of service. $\mathcal{B}$ then evaluates among all requests and offers it has receives a suitable assignments of clients' requests to available compute nodes. Alternatively, the clients and the compute nodes can directly discover and connect to each other. In such an approach, the clients serve as their own broker, rendering the clients heavyweight.  
Once $\mathcal{P}$ and $\mathcal{C}$ are matched, $\mathcal{P}$ commits a payment $v$ to an escrow on the Ethereum blockchain, and sends $\texttt{pkg} = \langle$\progKT, \encinp, \auxdata $\rangle$ to $\mathcal{C}$, wherein 
$\texttt{Enc}()$ is a symantically secure symmetric-key encryption scheme~\cite{crypto_intro}, and \auxdata\ contains auxiliary data that is needed for the execution (e.g., the proof that $\mathcal{P}$ has committed a payment $v$ to the escrow). $\mathcal{C}$ then instantiates a \progKT\ enclave, and attests that the enclave has been instantiated correctly. Upon successful attestation, a secret key $k_{\mathcal{P}}$ is provisioned to the \progKT\ enclave, allowing it to process and compute on \inp. Finally, the result of the computation, namely \output\ is sent to $\mathcal{P}$. We note that \output\ is encrypted in such a way that its decryption by $\mathcal{P}$ ensures full payment of $v$ to $\mathcal{C}$. Figure~\ref{fig:architecture} depicts a workflow and interactions between the three parties in \codename.

\subsection{System Goals}
\label{subsec:system_goals}
This section specifies primary desired properties that \codename\ aims to achieve, with respect to security, scalability, and usability. These properties motivate and justify \codename's design choices. 

\vspace{2mm}
\noindent\textbf{Data confidentiality:} \codename\ ensures that \inp, and secret states of \program\ from an honest client remain encrypted outside the enclave memory, and thus are not known to any other party, including the compute node. The key to decrypt them resides only inside the enclave. 
We remark that \codename\ does not attempt to eliminate \program's side-channel leakage. However, it is compatible with defense against these leakages~\cite{path_oram, oblivm, STC}. 

\vspace{2mm}
\noindent\textbf{Correct and attested execution:} \codename\ ensures that the output obtained by the client correctly reflects the faithful execution of \program\ on \inp.

\vspace{2mm}
\noindent\textbf{Fair exchange:} \codename\ assures that the work a compute node exhausts in executing the outsourced task is accurately metered and remunerated in fine granularity. At the same time, \codename\ ensures that a compute node gets the full reward for the outsourced computation if and only if  a client gets a correct result of such computation.

\vspace{2mm}
\noindent\textbf{Limited interaction between client and compute node:} \codename\ unburdens clients from constantly maintaining a connection with their assigned compute nodes prior to and during execution of the outsourced computations. 


\def\vd{-0.75}
\def\bd{0.}
\def\above{0.16cm}
\def\ld{-0.2}
\def\rd{0.2}
\def\pw{1.2 cm}
\def\width{3.5}

\begin{figure}[t]
\centering    
\resizebox{.4\textwidth}{!}{  
\begin{tikzpicture}[every text node part/.style={align=center}]

\node[rectangle, draw = black, thick](B){\large $\mathcal{B}$};
\node[rectangle, draw = black, right of = B, node distance = \width cm, thick](W){\large $\mathcal{C}$};
\node[rectangle, draw = black, left of = B, node distance = \width cm, thick](P){\large $\mathcal{P}$};
\node[ellipse, draw = black, thick, fill = green,  below of = B, node distance = -1*\vd cm](escrow) {Escrow};

\draw[->, thick] (P.east) -- node[yshift = \above] {(1a) Request}(B.west);
\draw[->, thick] (W.west) -- node[yshift = \above] {(1b) Resource offer}(B.east);
\draw[->, thick, dashed]  ($(P.south)+(0.15,0)$) -- ($(P.south)+(0.15,2*\vd)$) -| node[yshift = \above, xshift = - \width cm] {(2b) $\texttt{pkg} = \langle$\progKT, \encinp, \auxdata$\rangle$} ($(W.south)+(-0.15,0)$){};
\draw[<-, thick, dashed]  ($(P.south)+(-0.15,0)$) -- ($(P.south)+(-0.15,3*\vd)$) -| node[yshift = \above, xshift = - \width cm] {(4a) $\texttt{Enc}(k_{\mathcal{CP}}, \ \output)$} ($(W.south)+(0.15,0)$){};

\draw[<->, thick, dashed]  (P.north) -- ($(P.north)+(0,0.5)$) -| node[yshift = \above, xshift = - \width cm] {(3) \progKT\ enclave attestation $\&$ provision of $k_{\mathcal{P}}$} (W.north){};

\draw[->, thick] (P.south east) -- node[yshift = \above, rotate = -10, xshift = 0.35cm] {(2a) $v$}(escrow.west);
\draw[->, thick] (escrow.east) -- node[yshift = \above, rotate = 10, xshift = -0.25cm] {(4b) $v$}(W.south west);

\end{tikzpicture}
}
\caption{An overview of \codename\ architecture. The key $k_{\mathcal{CP}}$ is computed using a secret chosen by $\mathcal{C}$ and $\mathcal{P}$'s secret session key $k_{\mathcal{P}}$. Communication between $\mathcal{P}$ and $\mathcal{C}$ can be routed via $\mathcal{B}$. The payment escrow (depicted in greenly shaded ellippse) is secured by the blockchain.} \label{fig:architecture}
\end{figure}
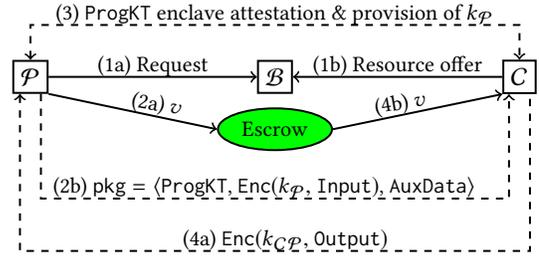
\subsection{Threat Model and Assumptions}
\label{subsec:threat_model}

We assume that an adversary is computationally bounded, and that cryptographic primitives employed in \codename\ are secure. We further assume that trusted processors provisioniong the TEE, in particular Intel SGX processor,  is implemented correctly and its protection mechanisms are not compromised. Although we do not consider side-channel attacks against the hardware~\cite{controlled_channel}, \codename\ is fully compatible with defenses against these attacks~\cite{oblivm, STC}. 

We do not consider denial of service attack wherein an adversary denies service to honest clients, or blocks honest compute nodes from the system. As such, \codename\ requires some compute nodes to behave correctly so as to guarantee the system's availability. We study a strong adversary that are capable of corrupting any number of clients and a majority (but not all) of the compute nodes. 

An adversary compromising a compute node can control its operating system, schedule its processes, reorder and tamper with its network messages. However, it cannot tamper with the enclaves' execution, nor observe theirs internal state. Compromised components can deviate arbitrarily from the prescribed protocol. 

Clients are mutually distrustful; i.e., they do not trust any other client. We assume that when a client delegates computation of the program \program, she trusts its code, and it is free of software bug. To enable fair exchange between clients and compute nodes, \codename\ necessitates instrumenting \program\ into \progKT\ that incorporates dynamic runtime checks over the execution of \program.  We assume that all parties can verify such instrumentation which is simple enough to lend itself to formal verification and vetting.

As mentioned earlier, the clients can serve as their own broker, handling compute node discovery, and performing their own request-resource matching. As such, there will always be honest, self-serving brokers in the system. In another word, \codename\ does not suffer from broker's single-point-of-failure problem.

\subsection{Design Roadmap}
\label{subsec:baseline}

This section introduces \baseline\ which serves as an elementary solution of a marketplace for outsourced computations. Subsequently, we analyze \baseline's shortcomings, based on which we outline design choices for \codename.

\paragraph{\baseline.} We assume that there exists a registry to which clients can post requests and compute nodes advertise their resources. The client $\mathcal{P}$ and compute node $\mathcal{C}$ actively scan for their counterpart. Once they find a match, $\mathcal{P}$ and $\mathcal{C}$ update the status of their request and offer to being served and occupied, respectively.

Let $v$ be the remuneration that $\mathcal{P}$ pays to $\mathcal{C}$ in exchange for executing \program\ on \inp\ and delivering the result \output. $\mathcal{P}$ sets up an escrow on the Ethereum blockchain with a deposit $v$, intended recipient $\mathcal{C}$, a time-out $T$, and a hash-lock $h$. This escrow holds the deposit $v$ until being invoked with a settling transaction \texttt{tx} containing \texttt{data} such that $H(\texttt{data}) = h$, wherein $H(\cdot)$ is a secure hash function~\cite{crypto_intro}. If \texttt{tx} arrives before the time-out $T$, the escrow sends $v$ to $\mathcal{C}$. Alternatively, if \texttt{tx} arrives after $T$ has expired, $v$ is refunded to $\mathcal{P}$.

\baseline\ provides a compiler that allows $\mathcal{P}$ to transform \program\ into a wrapper program \progSM. 
In addition to \inp, \progSM\ also consumes $h$ and \texttt{data}. It checks if \texttt{data} can indeed settle the escrow (i.e., $H(\texttt{data}) = h$) before executing \program\ on \inp\ to obtain \output. If \texttt{data} cannot settle the escrow, \progSM\ aborts the execution. We assume that $\mathcal{P}$ can formally verify the correctness of the compilation.

After setting up the escrow, $\mathcal{P}$ sends $\texttt{pkg} = \langle$\progSM, \encinp, \auxdata $\rangle$ to $\mathcal{C}$, wherein $\texttt{Enc}()$ is a symantically secure symmetric-key encryption scheme~\cite{crypto_intro}, and \auxdata\ contains $\texttt{Enc}(k_{\mathcal{P}}, \ \texttt{data})$ along with $h$. $\mathcal{C}$ then instantiates a \progSM\ enclave. Next, $\mathcal{P}$ and the enclave engage in a remote attestation procedure~\cite{sgx_remote_attest} which convinces  $\mathcal{P}$ that the code loaded within the enclave is indeed \progSM. Furthermore, the remote attestation  allows $\mathcal{P}$ to establish a secure authenticated channel with the enclave, via which they communicate the key $k_{\mathcal{P}}$. $\mathcal{C}$ supplies the enclave with \encinp\ and \auxdata. The enclave checks the validity of \texttt{data} before executing \program\ on \inp. Upon the completion of the computation, the \progSM\ returns to $\mathcal{C}$ the encrypted output $\texttt{Enc}(k_{\mathcal{P}}, \ \output)$ and \texttt{data}. $\mathcal{C}$ sends the encrypted output to $\mathcal{P}$, and claims the reward $v$ using \texttt{data}.

\vspace{2mm}
\noindent\textbf{\baseline's shortcomings.}
\baseline\ protects the integrity of the outsourced computation, the confidentiality of the client's data, and guarantees remuneration for the compute nodes once they finish the outsourced task. Nonetheless, \baseline\ still observes various shortcomings. 

First, \baseline\ rewards $\mathcal{C}$ based on task completion, which exposes the payment to various fairness issues. On the one hand, micro tasks yielding very small remunerations suffer from overhead incurred by the transaction fee. On the other hand, if $\mathcal{C}$ inadvertently aborts the computation midway due to its unexpectedly large resource consumption, it is not remunerated for the work it has completed prior to the abortion. Even worse, $\mathcal{C}$ could deny $\mathcal{P}$ of the computation result by dropping it after obtaining \texttt{data}, claiming the reward without delivering the encrypted output to $\mathcal{P}$.  

Second, \baseline\ requires $\mathcal{P}$ to remain online until its request is accepted by a compute node $\mathcal{C}$ to carry out a remote attestation procedure and provision the key $k_{\mathcal{P}}$ to the \progSM\ enclave. The procedure requires $\mathcal{P}$ to contact IAS for verifying the attestation it obtains from $\mathcal{C}$. This is clearly inconvenient for $\mathcal{P}$, especially when the request demands uncommon resources or the IAS service is temporarily unavailable.

Finally, \baseline\ assumes $\mathcal{P}$ and $\mathcal{C}$ could efficiently discover their counterparts. This assumption, however, may not hold true in practice, causing the system to be under-utilised when requests are not served despite there are available compute nodes. 

\section{\codename\ Design}
\label{sec:design}
This section introduces \codename's design to overcome \baseline's shortcomings. In particular,
\codename\ enables fair exchange between $\mathcal{P}$ and $\mathcal{C}$ via a trusted work metering mechanism coupled with a scalable protocol for micro payments. 
Further, it unburdens  $\mathcal{P}$ from remaining online and engaging in a remote attestation with $\mathcal{C}$. 
Finally, \codename\ entrusts $\mathcal{B}$ to match clients' requests with available compute nodes, maximizing the system's resource utilization.

\subsection{Fair Exchange}
\label{subsec:fair_payment}
Unlike \baseline\ rewarding $\mathcal{C}$ based on task completion, \codename\ splits the reward $v$ of the outsourced 
computation into two portions, namely $v_c = \alpha v$ and $v_d = (1-\alpha)v$, where $\alpha$ is a parameter set by the client $\mathcal{P}$, and agreed upon by $\mathcal{C}$. 
The first portion (i.e., $v_c$) remunerates $\mathcal{C}$ for its work on a fine-grained basis, while the second portion (i.e., $v_d$) rewards the delivery of the result. 

More specifically, $\mathcal{C}$ is  entitled to $v_c$ upon the completion of the outsourced computation. In case $\mathcal{C}$ inadvertently aborts the computation midway, it is still remunerated with a fraction of $v_c$ 
according to its progress prior to the suspension.
The remaining portion of $v$, namely $v_d$, is only payable to $\mathcal{C}$ when the computation output is delivered to $\mathcal{P}$. This discourages $\mathcal{C}$ from denying $\mathcal{P}$ of the result as it may in \baseline. Additional mechanism that disincentivises result withholding (e.g., requiring $\mathcal{C}$ to make a security deposit which is forfeited should they repeatedly abort the computation~\cite{decentralised_poker, amortizing_MPC})  can also be incorporated into \codename.

\subsubsection{TEE-based metering.} 
\label{subsubsec:smart_meterting}
To enable a fair exchange described above, \codename\ has to meter the compute node's work in a fine-grained and tamper-proof fashion. We follow REM~\cite{REM} in implementing a reliable metering logic inside the enclave. More specifically, \codename\ requires the client's program \program\ to be instrumented into a wrapper program \progKT. The wrapper program reserves the logic of the original program (i.e., it executes \program's logic on \inp), while keeping a counter of the number of instructions that has been executed. This is then used as a measurement of the compute node's effort.

\progKT\ maintains the instruction counter in a reserved register which is inaccessiable to any other process. 
To prevent a malicious \program\ from manipulating the instruction counter, \codename\ does not support \program\ that is multi-threaded or contains writeable code pages~\cite{REM, enclave_writer_guide}. When the \progKT\ enclave halts or exits, it returns a ``proof of work'' (i.e., the number of instruction executed) based on which \codename\ settles the payment of $v_c$ (or a fraction of it). We remark that the restriction of single-threaded \program\ is not necessary a severe limitation, for threading in SGX enclave is much different compared to that of legacy software~\cite{sgx_note}. In particular, one cannot create or destroy an SGX thread on the fly, and an SGX thread is mapped directly to a logical processor. Consequently, a typical SGX-compliant program (i.e., a program that inherently supports SGX-enclave execution) is often single-threaded. 

\vspace{1mm}
\noindent\textbf{On the choice of instruction counting.} One may argue that instructions are not the most accurate metric for CPU effort. Alternative metrics include CPU time and CPU cycles. Nevertheless, these metrics are subject to manipulation by the malicious OS. Even if they were not manipulated, they are incremented even when an enclave is swapped out~\cite{REM}. Consequently, we believe that instruction counting is the most appropriate method for securely measuring the compute node's effort using available tools in SGX.

\subsubsection{Micro Payments with Off-Chain Payment Channel}
One naive approach to settle the proof of work is for $\mathcal{C}$ to send it to $\mathcal{P}$, who then responds with a transaction paying a corresponding amount of reward to $\mathcal{C}$. This approach, however, does not ensure fairness in case $\mathcal{P}$ neglects her outsourced computation. Another approach is to have $\mathcal{P}$ commit a number of equally-valued micro transactions, each of which contains a fraction of $v_c$, to a payment escrow on the blockchain, and to structure the proof such that it can be used to autonomously claim a subset or all of those micro transactions. Nonetheless, settling a large number of micro transactions on the blockchain incurs high overhead. 

\codename\ sidesteps this challenge by leveraging payment channel~\cite{raiden_eth}, allowing two parties to transact a large number of micro payments without incurring high transaction fee or overloading the blockchain with excessive number of transactions. It is assumed that a payer and a payee maintain a payment channel (discussed in section~\ref{subsec:payment_channel}), and each micro payment is represented by a payment promise to be communicated off-chain (i.e., off the blockchain) between the payer and the payee. To settle the payments, the payee posted the latest payment promise it has received, claiming the sum of all promises and thereby closing the channel. However, establishing a new channel for each pair of client and compute node is inefficient. \codename, instead, makes use of multi-hop channels\footnote{While the channels can be bidirectional, our discussion focuses on unidirectional channels. Extending \codename\ to support bidirectional channels is trivial.} to better utilize the channel capacity, requiring fewer channels to be established. 

To this end, \codename\ assumes that each client $\mathcal{P}$  maintains a payment channel with the broker $\mathcal{B}$ that, in turn, maintains a channel with each compute node $\mathcal{C}$. 
The payment from $\mathcal{P}$ to $\mathcal{C}$ does not require the two parties to establish a channel. Instead, it could be securely routed via $\mathcal{B}$, in a sense that if $\mathcal{C}$ collects a payment from $\mathcal{B}$, it is guaranteed that $\mathcal{B}$ could also collect a corresponding payment from $\mathcal{P}$\footnote{$\mathcal{B}$ could charge $\mathcal{P}$ a service fee in return to routing the payment. Nonetheless, for simplicity, we assume $\mathcal{B}$ offers such routing free of charge. Extending \codename\ to support such service fee is trivial.}. 
We assume that each payment channel has sufficiently large capacity (i.e., the on-chain deposit that was set up at the beginning of the channel) to accommodate the payment of various outsourced computations during its lifetime. 
\def\vd{-0.6}
\def\bd{0.6}
\def\above{0.2cm}
\def\ld{-0.2}
\def\rd{0.2}
\def\pw{1.2 cm}
\def\width{5}

\begin{figure}[t]
\centering    
\resizebox{.48\textwidth}{!}{  
\begin{tikzpicture}[every text node part/.style={align=center}]

\node[rectangle, draw = black, thick](B){\large $\mathcal{B}$};
\node[rectangle, draw = black, right of = B, node distance = \width cm, thick](W){\large $\mathcal{C}$};
\node[rectangle, draw = black, left of = B, node distance = \width cm, thick](P){\large $\mathcal{P}$};

\node[rectangle, draw = none, below of = B, node distance = 12.8*\bd cm](BE){};
\node[rectangle, draw = none, right of = BE, node distance = \width cm](WE){};
\node[rectangle, draw = none, left of = BE, node distance = \width cm](PE){};

\draw[->, dashed] (P.south) -- (PE.south);
\draw[->, dashed] (B.south) -- (BE.south);
\draw[->, dashed] (W.south) -- (WE.south);

\node[below of = B, node distance = \bd cm, fill = white]{Pick $\texttt{rand}_{\mathcal{B}}$};

\draw[->, thick] ($(B.west) + (\rd,1.8*\vd)$) -- node[yshift = \above] {$h_{\mathcal{B}} = H(\texttt{rand}_{\mathcal{B}})$} ($(P.east) +(\ld,1.8*\vd)$);

\node[below of = P, node distance = 2.4*\bd cm, fill = white, xshift = 1.25cm]{Pick $\langle s_1, s_2, \ldots s_n \rangle$, compute $h_i = H(s_i)$};
\node[below of = P, node distance = 3.3*\bd cm, fill = white, xshift = 1.25cm]{Pick $\texttt{rand}_{\mathcal{P}}$, compute
 $h_{\mathcal{P}} = H(\texttt{rand}_{\mathcal{P}})$};

\node[rectangle, minimum width = 0.25cm, fill = black, draw = black, below of = P, node distance = 4.4*\bd cm, xshift = 0.5cm](lc10){};
\node[ellipse, minimum height = 0.5cm, scale = 0.8, draw = black, thick, below of = lc10, node distance = -0.05 cm](lc11){};
\node[rectangle, minimum width = 0.25cm, fill = white, below of = lc10, node distance = 0.22 cm](lc12){};
\node[below of = lc10, node distance = -0.2cm, xshift = 0.3cm]{$h_1$};
\node[rectangle, minimum width = 0.6cm, draw = black, below of = P, node distance = 4.85*\bd cm, xshift = 0.85cm, text width=0.55cm,align=right,](promisec1){$m^{\mathcal{B}}_1$};

\node[rectangle, draw = none, right of = promisec1, node distance = 0.8cm](dots1){\Large \textbf{.  .  .}};

\node[rectangle, minimum width = 0.25cm, fill = black, draw = black, below of = P, node distance = 4.4*\bd cm, xshift = 2.1cm](lcn10){};
\node[ellipse, minimum height = 0.5cm, scale = 0.8, draw = black, thick, below of = lcn10, node distance = -0.05 cm](lcn11){};
\node[rectangle, minimum width = 0.25cm, fill = white, below of = lcn10, node distance = 0.22 cm](lcn12){};
\node[below of = lcn10, node distance = -0.2 cm, xshift = 0.3cm]{$h_n$};
\node[rectangle, minimum width = 0.6cm, draw = black, below of = P, node distance = 4.85*\bd cm, xshift = 2.45cm](promisecn1){$m^{\mathcal{B}}_n$};

\node[rectangle, minimum width = 0.25cm, fill = black, draw = black, below of = P, node distance = 4.4*\bd cm, xshift = 3.3cm](l10){};
\node[ellipse, minimum height = 0.5cm, scale = 0.8,draw = black, thick, below of = l10, node distance = -0.05 cm](l11){};
\node[rectangle, minimum width = 0.25cm, fill = white, below of = l10, node distance = 0.22 cm](l12){};
\node[rectangle, minimum width = 0.25cm, fill = black, draw = black, right of = l10, node distance = \pw](l20){};
\node[ellipse, minimum height = 0.5cm, scale = 0.8, draw = black, thick, right of = l11, node distance = 1.5cm](l21){};
\node[rectangle, minimum width = 0.25cm, fill = white, right of = l12, node distance = \pw](l22){};
\node[below of = l10, node distance = -0.2 cm, xshift = 0.3cm]{$h_{\mathcal{P}}$};
\node[below of = l20, node distance = -0.2 cm, xshift = -0.3cm]{$h_{\mathcal{B}}$};
\node[rectangle, minimum width = \pw, draw = black, below of = P, node distance = 4.85*\bd cm, xshift = 3.9cm](promise1){$m^{\mathcal{B}}_d$};

\draw[->, thick] ($(P.east) + (\ld,5.75*\vd)$) -- node[yshift = \above] {} ($(B.west) +(\rd,5.75*\vd)$);

\node[below of = W, node distance = 6.25*\bd cm, fill = white]{Pick $\texttt{rand}_{\mathcal{C}}$};
\draw[->, thick] ($(W.west) + (\rd ,7*\vd)$) -- node[yshift = \above] {$h_{\mathcal{C}} = H(\texttt{rand}_{\mathcal{C}})$} ($(B.east) +(\ld,7*\vd)$);

\node[rectangle, minimum width = 0.25cm, fill = black, draw = black, below of = B, node distance = 8.2*\bd cm, xshift = 0.5cm](lc20){};
\node[ellipse, minimum height = 0.5cm, scale = 0.8, draw = black, thick, below of = lc20, node distance = -0.05 cm](lc21){};
\node[rectangle, minimum width = 0.25cm, fill = white, below of = lc20, node distance = 0.22 cm](lc22){};
\node[below of = lc20, node distance = -0.2cm, xshift = 0.3cm]{$h_1$};
\node[rectangle, minimum width = 0.6cm, draw = black, below of = B, node distance = 8.65*\bd cm, xshift = 0.85cm, text width=0.55cm,align=right,](promisec2){$m^{\mathcal{C}}_1$};

\node[rectangle, draw = none, right of = promisec2, node distance = 0.8cm](dots2){\Large \textbf{.  .  .}};

\node[rectangle, minimum width = 0.25cm, fill = black, draw = black, below of = B, node distance = 8.2*\bd cm, xshift = 2.1cm](lcn20){};
\node[ellipse, minimum height = 0.5cm, scale = 0.8, draw = black, thick, below of = lcn20, node distance = -0.05 cm](lcn21){};
\node[rectangle, minimum width = 0.25cm, fill = white, below of = lcn20, node distance = 0.22 cm](lcn22){};
\node[below of = lcn20, node distance = -0.2 cm, xshift = 0.3cm]{$h_n$};
\node[rectangle, minimum width = 0.6cm, draw = black, below of = B, node distance = 8.65*\bd cm, xshift = 2.45cm](promisecn2){$m^{\mathcal{C}}_n$};

\node[rectangle, minimum width = 0.25cm, fill = black, draw = black, below of = B, node distance = 8.2*\bd cm, xshift = 3.3cm](l60){};
\node[ellipse, minimum height = 0.5cm, scale = 0.8,draw = black, thick, below of = l60, node distance = -0.05 cm](l61){};
\node[rectangle, minimum width = 0.25cm, fill = white, below of = l60, node distance = 0.22 cm](l62){};
\node[rectangle, minimum width = 0.25cm, fill = black, draw = black, right of = l60, node distance = \pw](l63){};
\node[ellipse, minimum height = 0.5cm, scale = 0.8, draw = black, thick, right of = l61, node distance = 1.5cm](l64){};
\node[rectangle, minimum width = 0.25cm, fill = white, right of = l62, node distance = \pw](l65){};
\node[below of = l60, node distance = -0.2 cm, xshift = 0.3cm]{$h_{\mathcal{P}}$};
\node[below of = l63, node distance = -0.2 cm, xshift = -0.35cm]{$h_{\mathcal{C}}$};
\node[rectangle, minimum width = \pw, draw = black, below of = B, node distance =8.65*\bd cm, xshift = 3.9cm](promise2){$m^{\mathcal{C}}_d$};

\draw[->, thick] ($(B.east) + (\ld,9.5*\vd)$) -- node[yshift = \above] {} ($(W.west) +(\rd ,9.5*\vd)$);

\node[below of = W, node distance = 10.25*\bd cm, fill = white, xshift = -1.25cm]{Collect $\texttt{Enc}(k_{\mathcal{CP}}, \ \output)$  from \progKT};

\draw[->, thick] ($(W.west) + (\rd,11.5*\vd)$) -- node[yshift = \above,xshift = -0.025cm] {$\texttt{Enc}(k_{\mathcal{CP}}, \ \output)$} ($(P.east) +(\ld,11.5*\vd)$);

\draw[->, thick] ($(P.east) + (\ld,12.35*\vd)$) -- node[yshift = \above, xshift = -0.01cm] {$\texttt{rand}_{\mathcal{P}}$} ($(W.west) +(\rd,12.35*\vd)$);

\end{tikzpicture}
}
\caption{An overview of the fair exchange in \codename. The payment promises $m^{\mathcal{B}}_i$ and $m^{\mathcal{C}}_i$ are hash-locked by $h_i$. The payment promise $m^{\mathcal{B}}_d$ is hash-locked by $h_{\mathcal{P}}$ and $h_{\mathcal{B}}$, while $m^{\mathcal{C}}_d$ is hash-locked by $h_{\mathcal{P}}$ and $h_{\mathcal{C}}$. The key $k_{\mathcal{CP}}$ is computed from  $\texttt{rand}_{\mathcal{C}}$ and the secret key $k_{\mathcal{P}}$.} \label{fig:fair_output_delivery}
\end{figure}
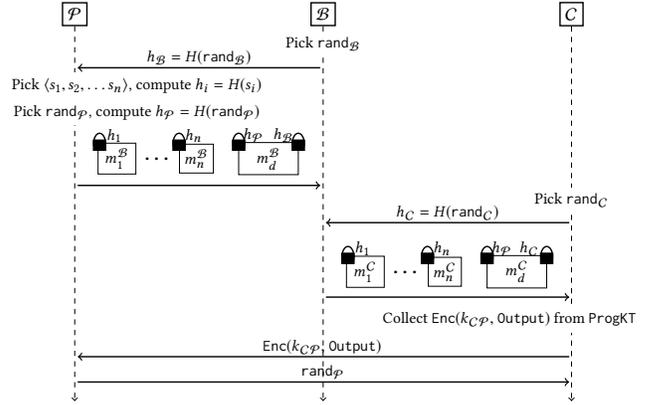

Figure~\ref{fig:fair_output_delivery} summarizes the fair exchange of the reward $v$ and the outsourced computation of \progKT.
The payment of $v$ is split over $n+1$ micro payments, $n$ of which summing up to $v_c$, while  the last one is worth $v_d$. 
The protocol does not require any communication between $\mathcal{P}$ and $\mathcal{C}$ \textit{prior to or during} the computation, nor a direct payment channel between the two parties. It, however, requires an off-chain communication between $\mathcal{P}$ and $\mathcal{C}$ in the final step to decrypt the output.

\vspace{0.2cm}
\noindent\textbf{Payment of $v_c$.}  Without loss of generality, let us assume that the payment of $v_c$ is divided into $n$ equally-valued payment promises, which are routed via $\mathcal{B}$. That is, $\mathcal{P}$ generates $n$ payment promises to $\mathcal{B}$, and $\mathcal{B}$ generates the corresponding $n$ payments promises to $\mathcal{C}$ with the same value and claiming condition. 

To generate the $n$ payment promises $\langle m^{\tiny{\mathcal{B}}}_1, m^{\mathcal{B}}_2, \ldots m^{\mathcal{B}}_n \rangle$ to $\mathcal{B}$, $\mathcal{P}$ first picks $n$ random strings $\langle s_1, s_2, \ldots s_n \rangle$, and computes their hashes $\langle h_1, h_2, \ldots h_n \rangle$ (i.e.,  $h_i = H(s_i)$).  
A digest $h_i$ is used to lock a promise $m^{\mathcal{B}}_i$, such that $\mathcal{B}$ can only use $m^{\mathcal{B}}_i$ to close the channel if it is aware of $\texttt{s}_i$ such that $H(\texttt{s}_i) = h_i$. 
The payment promise $m^{\mathcal{B}}_i$ is worth $[\texttt{debt}_{\mathcal{P}} + (i \times v_c)/n]$ wherein $
\texttt{debt}_{\mathcal{P}}$ is the accumulated amount of unsettled payment for $\mathcal{P}$'s previous requests. 
Finally, $\mathcal{P}$ encrypts the random strings $\langle s_1, s_2, \ldots s_n \rangle$ with $k_{\mathcal{P}}$, and attaches them as well as the payment promises to \auxdata.

Similarly, $\mathcal{B}$ generates the corresponding promises 
$\langle m^{\mathcal{C}}_1, m^{\mathcal{C}}_2, \ldots m^{\mathcal{C}}_n \rangle$ to $\mathcal{C}$. Each promise $m^{\mathcal{C}}_i$ is locked by $h_i$ (i.e., the same hash-lock as $m^{\mathcal{B}}_i$), and worth $[\texttt{cred}_{\mathcal{C}} + (i \times v_c)/n]$ wherein $\texttt{cred}_{\mathcal{C}}$ is the accumulated unsettled credit that $\mathcal{C}$ is entitled to claim for its previous services. $\mathcal{B}$ includes these promises into the  \auxdata\ before forwarding \texttt{pkg} to $\mathcal{C}$.

\vspace{0.2cm}
\noindent\textbf{Payment of $v_d$ upon output delivery.}
To ensure that the remaining portion of $v$, namely $v_d$, can only be collected upon the delivery of the output to $\mathcal{P}$, \progKT\ encrypts the output  using a key $k_{\mathcal{CP}}$ derived from $k_{\mathcal{P}}$ and a secret $\texttt{rand}_{\mathcal{C}}$ chosen and committed to by ${\mathcal{C}}$. At the same time, the full payment of $v$ is encumbered until the disclosure of $\texttt{rand}_{\mathcal{C}}$. 

As shown in Figure~\ref{fig:fair_output_delivery}, besides the $n$ payment promises above, $\mathcal{P}$ generates another payment promise $m^{\mathcal{B}}_d$ to $\mathcal{B}$ that is worth $[\texttt{debt}_{\mathcal{P}} + v]$ and is hash-locked by two digests $h_{\mathcal{B}}$ and $h_{\mathcal{P}}$.
Similarly, $\mathcal{B}$ also generate one more payment promise $m^{\mathcal{C}}_d$ to $\mathcal{C}$ that is worth $[\texttt{cred}_{\mathcal{C}} + v]$, 
and hash-locked by $h_{\mathcal{P}}$ and $h_{\mathcal{C}}$. The three hash-locks $h_{\mathcal{P}}$, $h_{\mathcal{B}}$ and $h_{\mathcal{C}}$ can be settled 
by three independent settling-data $\texttt{rand}_{\mathcal{P}}$, $\texttt{rand}_{\mathcal{B}}$ and $\texttt{rand}_{\mathcal{C}}$ chosen independently at random by the three parties ${\mathcal{P}}, {\mathcal{B}}$ and ${\mathcal{C}}$, respectively.

\vspace{0.2cm}
\noindent\textbf{Dynamic Runtime Checks.}
The fair exchange requires the wrapper enclave \progKT\ to perform some dynamic checks at runtime prior to executing \program's logic. More specifically, besides \inp\ and \auxdata, \progKT\ also consumes the  hash-lock $h_{\mathcal{C}}$ and $\texttt{rand}_{\mathcal{C}}$. It first verifies the validity of the settling-data $\langle s_1, s_2, \ldots s_n \rangle$ (i.e., $h_i = H(s_i) \forall \langle h_i, s_i \rangle \in$ \auxdata). Next, it checks if $h_{\mathcal{C}} = H(\texttt{rand}_{\mathcal{C}})$. Only when the verification passes does it execute \program\ on \inp, obtaining \output. It then encrypts \output\ with $k_{\mathcal{CP}} = k_{\mathcal{P}} \oplus \texttt{rand}_{\mathcal{C}}$, producing an encrypted output $\texttt{Enc}(k_{\mathcal{CP}}, \ \output)$. Finally, the enclave returns the appropriate settling-data $s_i$ based on the instruction counter and the encrypted output (if it successfully completes the computation) to $\mathcal{C}$.

\vspace{0.2cm}
\noindent\textbf{Payment settlement.}
The  settling-data $s_i$ renders the promise $m^{\mathcal{C}}_i$ claimable, enabling $\mathcal{C}$ to collect (a portion of)  $v_c$ according to its work. To obtain the settling-data necessary to claim $m^{\mathcal{C}}_d$ (i.e., the full reward $v$), $\mathcal{C}$ has to 
send the encrypted output to $\mathcal{P}$, who then responds with $\texttt{rand}_{\mathcal{P}}$. If $\mathcal{C}$ chooses to settle the payment thereby closes the channel between $\mathcal{C}$ and $\mathcal{B}$, it has to reveal both $\texttt{rand}_{\mathcal{P}}$ and $\texttt{rand}_{\mathcal{C}}$. 
This allows $\mathcal{P}$ to compute $k_{\mathcal{CP}}$ and obtain \output, and $\mathcal{B}$ to claim $m^{\mathcal{B}}_d$. Alternatively, should $\mathcal{C}$ wish to maintain the channel, it has to back propagate the settling-data to $\mathcal{B}$ and $\mathcal{P}$ so that they can update $\texttt{cred}_{\mathcal{C}}$ and  $\texttt{debt}_{\mathcal{P}}$ accordingly\footnote{In case the computation is inadvertently aborted midway and thus no output is 
produced, the promise to be settled is $m^{\mathcal{C}}_i$ and settling-data is $s_i$.}. In a situation where $\mathcal{P}$'s response is 
invalid (i.e., its digest produced by the standard hash function $H(\cdot)$ does not match $h_{\mathcal{P}}$), $\mathcal{C}$ can check this invalidity locally and use it as a evidence to accuse $\mathcal{P}$ of conducting mischief. In such situation, 
fairness property is still guaranteed; i.e., $\mathcal{C}$ does not claim $v_d$ from $\mathcal{B}$, who in turn does not claim  $v_d$ from $\mathcal{P}$ and $\mathcal{P}$ cannot decrypt $\texttt{Enc}(k_{\mathcal{CP}}, \ \output)$ to obtain \output.

\subsection{Delegated Attestation}
\label{subsec:delegate_attest}
Unlike \baseline, \codename\ relieves $\mathcal{P}$ from conducting a remote attestation with $\mathcal{C}$ at the beginning of every request execution by implementing a  \textit{delegated attestation} scheme. The scheme requires each broker $\mathcal{B}$ to run an attestation manager enclave \BAM \ (detailed in Algorithm~\ref{alg:BAM}), and each compute node $\mathcal{C}$ to run a key handler enclave \WKH\ (detailed in Algorithm~\ref{alg:WKH}). The execution of \BAM\ and \WKH\ are protected by Intel SGX. 

Without loss of generality, the delegated attestation builds a chain of trust that comprises three links. The first and second links are established via remote attestations between $\mathcal{P}$ as a validator and  \BAM\ as an attesting enclave, and \BAM\ as a validator and  \WKH\ as an attesting enclave. The final link entails \progKT\ enclave to prove its correctness to \WKH\ via local attestation. Chaining all three links together, $\mathcal{P}$ gains confidence that the \progKT\ enclave has been properly instantiated on the compute node $\mathcal{C}$ using the correct code, without contacting  $\mathcal{C}$ or the IAS.  

\begin{algorithm}[t]
\caption{Attestation Manager Enclave}
\label{alg:BAM}
\begin{algorithmic}
\Procedure{ReceiveKeyFrom$\mathcal{P}$}{$k_{\mathcal{P}}$} 
\State keyID $\gets$ Seal$(k_{\mathcal{P}})$; 
\State \textcolor{gray}{\footnotesize // encrypts $k_{\mathcal{P}}$ with the enclave's Seal Key for persistent storage to disk}
\State \Return keyID
\EndProcedure
\Procedure{VerifyCert}{\certW, ${M}_{\texttt{KH}}$}
\State $b_1 \gets $VerifySignature (\certW);
\State $\pi_{\texttt{KH}} \gets$ GetAttestation(\certW);
\State $b_2 \gets $CheckMeasurement$(\pi_{\texttt{KH}}, {M}_{\texttt{KH}})$; 
\State \Return $b_1 \wedge b_2 $
\EndProcedure
\Procedure{ProvisionKey}{keyID, \certW}
\State $k_{\mathcal{P}} \gets$ UnSeal(keyID); 
\State $\texttt{pk}_{\texttt{KH}} \gets $ GetPK(\certW);
\State $\texttt{channelID} \gets $ EstablishSecureChannel($\mathcal{C}, \texttt{pk}_{\texttt{KH}}$);
\State \textcolor{gray}{\footnotesize // inter-platform channel between $\mathcal{B}$'s \BAM\ and $\mathcal{C}$'s \WKH}
\State Send($\texttt{channelID}, k_{\mathcal{P}}$); 
\EndProcedure
\end{algorithmic}
\end{algorithm}

Each attestation manager enclave has its own (unique) public and private key pair ( $\texttt{pk}_{\texttt{AM}}$, $\texttt{sk}_{\texttt{AM}}$) that are generated uniformly at random during the enclave instantiation.  Upon successfully instantiating \BAM, $\mathcal{B}$ requests the trusted processor for its remote attestation $\pi_{\texttt{AM}} = \langle {M}_{\texttt{AM}}, \texttt{pk}_{\texttt{\texttt{AM}}} \rangle_{\sigma_{TEE}}$, where ${M}_{\texttt{AM}}$ is the enclave's measurement, and $\sigma_{TEE}$ is a group signature signed by the processor's private key. The certificate $\pi_{\texttt{AM}}$ attests for the correctness of the \BAM\ enclave and its public key. Nonetheless, the only party that can verify $\pi_{\texttt{AM}}$ is the IAS acting as group manager~\cite{sgx_remote_attest}. \codename\ converts  $\pi_{\texttt{AM}}$ into a \textit{publicly verifiable} certificate by having  $\mathcal{B}$ obtain and store the IAS response  \certB\ = $\langle \pi_{\texttt{AM}}, \texttt{validity} \rangle_{\sigma_{IAS}}$ where $\sigma_{IAS}$ is the IAS's publicly verifiable signature on $\pi_{\texttt{AM}}$ and the $\texttt{validity}$ flag. By examining \certB, the enclave code and its measurement ${M}_{\texttt{AM}}$, any party can verify the correctness of and establish a secure connection to the \BAM\ enclave. 

Likewise, every compute node $\mathcal{C}$ runs a key handler enclave \WKH\ .  $\mathcal{C}$ obtains (from the IAS) and stores a publicly verifiable certificate \certW = $\langle \pi_{KH}, \texttt{valid} \rangle_{\sigma_{IAS}}$, 
where $\pi_{KH}$ is \WKH's remote attestation containing its measurement ${M}_{KH}$ and its unique public key $\texttt{pk}_{\texttt{KH}}$. By examining \certW, any party can be assured of the correctness of \WKH\ and communicate securely with it.

\begin{algorithm}[t]
\caption{Key Handler Enclave}
\label{alg:WKH}
\begin{algorithmic}
\Procedure{ReceiveKeyFrom\BAM}{$k_{\mathcal{P}}$}
\State keyID $\gets$ Seal$(k_{\mathcal{P}})$; 
\State \textcolor{gray}{\footnotesize // encrypts $k_{\mathcal{P}}$ with the enclave's Seal Key for persistent storage to disk}
\State \Return keyID
\EndProcedure
\Procedure{VerifyLocalAtt}{$\psi_{\texttt{ProgKT}}$, ${M}_{\texttt{ProgKT}}$}
\State \textcolor{gray}{\footnotesize // $\psi_{\texttt{prog}}$ is local attestation of the \progKT\ enclave}
\State $b_1 \gets $VerifyMAC ($\psi_{\texttt{ProgKT}}$);
\State $b_2 \gets $CheckMeasurement$(\psi_{\texttt{ProgKT}}, {M}_{\texttt{ProgKT}})$; 
\State \Return $b_1 \wedge b_2 $
\EndProcedure
\Procedure{SendKeyLocal}{keyID, $\psi_{\texttt{prog}}$}
\State $k_{\mathcal{P}} \gets$ UnSeal(keyID); 
\State $\texttt{channelID} \gets $ EstablishLocalSecureChannel($\psi_{\texttt{ProgKT}}$);
\State \textcolor{gray}{\footnotesize // intra-platform channel, for both \WKH\ and \progKT\ are instantiated on $\mathcal{C}$}
\State Send($\texttt{channelID}, k_{\mathcal{P}}$); 
\EndProcedure
\end{algorithmic}
\end{algorithm}

\def\vd{-0.45}
\def\above{0.18cm}
\def\width{2.75}
\begin{figure}[t]
\centering    
\resizebox{.45\textwidth}{!}{  
\begin{tikzpicture}

\node[rectangle, draw = black](P){$\mathcal{P}$};
\node[rectangle, draw = black, right of = P, node distance = 3.5 cm, minimum width = 1cm, minimum height = 0.75cm, text width=1.0cm,align=right](B) {$\mathcal{B}$};
\node[rectangle, draw = black, right of = P, node distance = 3.28 cm, fill = green ](AM){\BAM};

\node[rectangle, draw = black, below of = P, node distance = \width cm, minimum width = 5.5cm, minimum height = 1cm, text width=5.5cm,align=right, xshift = 1.8 cm](W) {$\mathcal{C}$};

\node[rectangle, draw = black, left of = W, node distance = 2.25 cm, fill = green](KH){\WKH};
\node[rectangle, draw = black, right of = KH, node distance = 3.8 cm, fill = green](progkt){\progKT};

\draw[<-](P.east) -- node[yshift = \above] {(1) \certB} (B.west);

\draw[->]  (P.north) -- ($(P.north)+(0,-0.6*\vd)$) -| node[yshift = \above, xshift = -1.75 cm] {(2b) $k_{\mathcal{P}}$} (AM.north){};
\draw[->]  ($(P.north)+(-0.15,0)$) -- ($(P.south)+(-0.15,-2.75*\vd)$) -| node[yshift = \above, xshift = -0.75 cm] {(2b) $\texttt{pkg}$} ($(B.north)+(0.15,0)$){};

\draw[->]($(W.north)+(1.5,0)$) -- node[xshift = -1*\above, yshift = -0.1cm, rotate = 90] {(3) \certW} (AM.south);
\draw[->](AM.south west) -- node[yshift = \above, rotate = 35] {(4a) $k_{\mathcal{P}}$} (KH.north);
\draw[->]($(B.south)+(0.1,0)$) -- node[xshift = 1*\above, rotate = -90] {(4b) $\texttt{pkg}$} ($(W.north)+(1.82,0)$);

\node[draw = none, right of = W, node distance = 3.5cm, yshift = 0.9cm, text width=3cm, align=center](){(5) $\mathcal{C}$ instantiates \\ \progKT\ enclave};

\draw[<-]($(KH.east)+(0,0.1)$) -- node[yshift = 1*\above] {(6) $\psi_{{\texttt{Prog}}_{KT}}$} ($(progkt.west)+(0,0.1)$);
\draw[->]($(KH.east)+(0,-0.1)$) -- node[yshift = -1*\above] {(7) $k_{\mathcal{P}}$} ($(progkt.west)+(0,-0.1)$);

\end{tikzpicture}
}
\caption{An overview of the delegated attestation scheme. The trusted enclaves are depicted in greenly shaded rectanlges.} \label{fig:delegated_attest}
\end{figure}
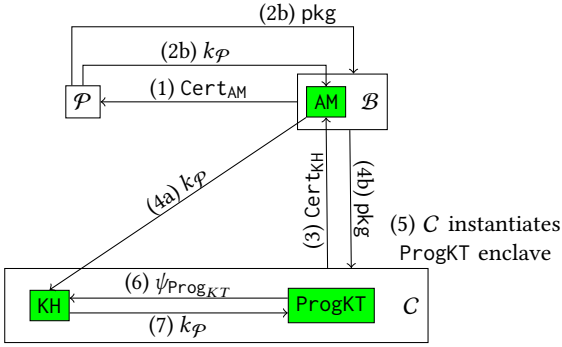

\vspace{0.2cm}
\noindent\textbf{Delegated Attestation Protocol.}
Figure~\ref{fig:delegated_attest} depicts the workflow of \codename's delegated attestation. 
After instrumenting \program\ into \progKT\ and verifying the correctness of the instrumentation, 
$\mathcal{P}$ initiates the delegated attestation by obtaining \certB\ from $\mathcal{B}$ and verifies its validity. It then establishes a secure and authenticated channel with \BAM\ using $\texttt{pk}_{\texttt{AM}}$. $\mathcal{P}$ then 
sends $\texttt{pkg} = \langle$\progKT\footnote{Should $\mathcal{P}$ want to hide \progKT\ from $\mathcal{B}$, she could also 
encrypt it with $k_{\mathcal{P}}$, and sends \encprogKT\ instead.}, \encinp, \auxdata$\rangle$ to $\mathcal{B}$,
and $k_{\mathcal{P}}$ to \BAM\ via the secure channel. Once $\mathcal{B}$ finds a compute node $\mathcal{C}$ that is willing to match $\mathcal{P}$'s request, \BAM\ obtains \certW\ from $\mathcal{C}$, verifies its validity, and establishes a secure and authenticated connection with $\mathcal{C}$'s \WKH\ to communicate $k_{\mathcal{P}}$. $\mathcal{B}$ then sends $\texttt{pkg}$ to $\mathcal{C}$. The compute node instantiates an enclave to execute \progKT, and performs a \textit{local attestation} with \WKH\ to prove its correctness. Upon successfully attestation, \WKH\ sends the key $k_{\mathcal{P}}$ to the \progKT\ enclave. Once the \progKT\ enclave completes the computation, it returns the encrypted output, which is then sent to $\mathcal{P}$ (perhaps being routed through $\mathcal{B}$). 

This mechanism only invokes IAS to obtain attestation certificates for \BAM\ and \WKH, instead of constantly involving IAS in every task execution as \baseline\ does. Further, it allows $\mathcal{P}$ to post a request (along with the payment) and then go offline until the time she wishes to collect the output, as opposed to remaining online till her request is picked up by some computation node.

\subsection{Task Assignment for Maximum Resource Utilization}
\label{subsec:task_assignment}
\codename\ designates to $\mathcal{B}$ the task of node discovery, load balancing and request-resource matching. More specifically, $\mathcal{B}$ collects requests from clients and advertise of available resource capacities from  compute nodes, and to evaluate the optimal assignments of requests to compute nodes that maximises the resource utilization of compute nodes or number of requests that are served at every time instance.

We assume that each request is associated with a resource specification requirement, and each resource offer indicates the compute node's available computational capacity. We further assume that there is no need for redundant execution (i.e., each request is 
assigned to one compute node), and a compute node serves only one request at a time. Let  $R = \langle r_1, r_2, \ldots, r_p \rangle$ be the set of pending requests, and $O = \langle {\mathcal{C}}_1, {\mathcal{C}}_2, \ldots, {\mathcal{C}}_q \rangle$ be the set of available compute 
nodes. The request assignment of $R$ to $O$ is the set $A$ comprising tuples of form $\langle r_i, {\mathcal{C}}_j \rangle$, wherein a compute node ${\mathcal{C}}_j$ has sufficient computational capacity to serve a request $r_i$. The optimization goal is to maximize the number of such tuples (i.e., $|A|$).

The request assignment problem is reducible to a \textit{maximum bipartite matching} problem~\cite{graph_theory}. 
A bipartite graph consists of vertices that can be divided into two independent sets such that there is no edge connecting vertices of same set. A maximum matching in a bipartite graph is a largest subset of the graph's edges such that no pair of edges in such a subset share an endpoint. 

To reduce \codename's request  assignment problem to the maximum bipartite matching problem, we create a graph $G=(V,E)$ wherein $V$ is the set of vertices, and $E$ is the set of edges. $V$ contains two independent subsets, $V_1$ and $V_2$, wherein $|V_1| = |R|$ and $|V_2| = |O|$. We represent each request $r_i$ by a vertex $v_i \in V_1$, and a compute node ${\mathcal{C}}_j$ by a 
vertex $v_{|R|+j} \in V_2$. We add an edge $e_{ij}$ that connects $v_i$ and $v_{|R|+j}$ if the compute node ${\mathcal{C}}_j$ has sufficient computational capacity to satisfy the resource specification requirement of request $r_i$. Since every edge in the graph $G$ connects a vertex in $V_1$ to a vertex in $V_2$, $G$ is a bipartite graph. 
The maximum matching in $G$ indicates the maximum request assignment of $R$ to $O$, in which an edge $e_{uv}$ in the maximum matching suggests the request $r_u$ to be assigned to the compute node ${\mathcal{C}}_v$.

We expect that a typical request in \codename\ can be served by a common compute node. This assumption renders the bipartite graph that represents all potential assignments of requests to compute nodes dense. The most efficient algorithm to find a maximum matching in a dense bipartite graph is the Mucha-Sankowski algorithm~\cite{max_matching}. This algorithm is randomized, and operates based on the fast matrix multiplcation algorithm. As such, it attains a running time complexity of $O(|V|^\omega$), where  $\omega$ is the exponent of the best known matrix multiplication algorithm (the best known algorithm to date is Le Gall algorithm~\cite{fast_matrix_mult}, with $\omega < 2.373$).

 \section{Security Analysis}
\label{sec:security_analysis}

\subsection{Fair exchange}
\label{subsec:payment_sec}
\noindent\textbf{TEE-based metering.}
To enable fair exchange between client's payment and compute node's computation, \codename\ necessitates dynamic runtime checks incorporated within the enclave that houses the outsourced computation. We implement this by providing a compiler that instruments any SGX-compliant program \program\ into a wrapper program \progKT. We believe that these additional steps and the overall instrumentation are simple enough to lend themselves to formal verification and vetting by \program\ writer, or by the client. 

As we mentioned earlier, the original \program\ should not contain writable code pages, for they would allow the program to rewrite itself at runtime and thus evade the instrumentation. This could be enforced by requiring the code page to have either {\em write} or {\em executable} permission exclusively (i.e., it cannot have both permission at the same time). This practice has also been recommended by Intel to the enclave writers~\cite{enclave_writer_guide}. 

In addition, \codename\ requires \program\ to be single-threaded. While the instruction counter is maintained in a reserved register which is inaccessible to any other processes (Section \ref{subsubsec:smart_meterting}), it remains accessible by different threads of \program, should it be multi-threaded. Thus, a malicious program that has multiple threads could manipulate the instruction counter value by carefully crafting the interactions of its threads. 

\vspace{0.2cm}
\noindent\textbf{Payment of $v_c$.}
\codename\ builds on payment channel~\cite{raiden_eth} to enable efficient micro payments and relies on the security of the Ethereum blockchain to ensure payment escrow is faithfully executed. To optimize for efficiency and avoid overloading the blockchain, \codename\ securely routes payment from $\mathcal{P}$ to $\mathcal{C}$ via the broker $\mathcal{B}$. A careful design of hash-lock payment promises, wherein promise from $\mathcal{P}$ to $\mathcal{B}$, and that of $\mathcal{B}$ to $\mathcal{C}$ could be settled using the same settling-data, guarantees that $\mathcal{B}$ can always claim from $\mathcal{P}$ which he pays to $\mathcal{C}$ on behalf of $\mathcal{P}$.  

\vspace{0.2cm}
\noindent\textbf{Ensuring Output Delivery.}
At the end of the computation, \progKT\ enclave encrypts the \output\ using key $k_{\mathcal{CP}} = k_{\mathcal{P}} \oplus \texttt{rand}_{\mathcal{C}}$. Since $m^{\mathcal{C}}_d$ is partially locked by $\texttt{rand}_{\mathcal{C}}$, the decryption of the output and the settling of $m^{\mathcal{C}}_d$ are bound together. Should $\mathcal{C}$ deny $\mathcal{P}$ of the output, it would have to forfeit $v_d$. While this act weakens the availability of the system, it does not violate fairness guarantee.

\subsection{Delegated Attestation}
\codename's delegated attestation relies on \BAM\ and \WKH\ enclaves to attest correct instantiation of \progKT\ enclave. Therefore, their correct instantiations are of utter importance. Fortunately, these enclave are fixed (as opposed to the \progKT\ enclave that houses client-defined program), and thus are easy to vet and verify. 

\codename's delegated attestation requires minimal involvement of $\mathcal{P}$ (i.e., examine the publicly verifiable certificates \certB\ = $\langle \pi_{\texttt{AM}}, \texttt{validity} \rangle_{\sigma_{IAS}}$). 
By checking that $\pi_{\texttt{AM}}$ indeed contains the expected measurement ${\mathcal{M}}_{\texttt{AM}}$, that its validity flag indicates \texttt{valid}, and that the certificate has been properly certified (using Intel's published public key~\cite{ias_pk}), $\mathcal{P}$ can ascertain the correct instantiation of  \BAM.  Moreover, using the public key $\texttt{pk}_{\texttt{\texttt{AM}}}$ included in $\pi_{\texttt{AM}}$, $\mathcal{P}$ can establish a secure and authenticated channel to \BAM\ via which the secret key $k_{\mathcal{P}}$ is communicated. Likewise, \BAM\ can verify the correct instantiation of \WKH\, and securely communicate $k_{\mathcal{P}}$ to the latter in the exact same manner. The security of the local attestation and communication between \WKH\ and \progKT\ enclave follows directly from Intel SGX's specifications~\cite{sgx_remote_attest}. Therefore, provided that cryptographic primitives in use are secure, and SGX hardware protection mechanisms are not subverted, \codename's delegated attestation is secure.

\subsection{Attested Execution and Data Confidentiality}
\label{subsec:sgx_sec}
\codename's relies on Intel SGX~\cite{sgx} to offer attested execution and data confidentiality to outsourced computations. In particular, SGX enables isolated execution~\cite{isolated_execution} ensuring that code loaded and running inside the enclaves cannot be tampered with by any other processes including the operating system or hypervisor. This, in combination with attestation capabilities, allows \codename\ to offer attested execution in which the computation correctness is guaranteed. Moreover, data  (i.e., input, output) and secret states of the enclave execution always remain encrypted outside of the enclave memory, thus their confidentiality are guaranteed. Furthermore, SGX memory encryption engine is capable of protecting data integrity and preventing memory replay attacks~\cite{prevent_replay_attack, rote}. 

Nonetheless, SGX's attested execution does not inherently offer protections against side-channel leakages~\cite{controlled_channel, m2r, page_fault_sgx}. The access pattern incurred by data (or code page) moving between the enclave and the non-enclave environment (e.g., page fault) could leak sensitive information about the code or data being processed within the enclave. Such side-channel leakage could be mitigated by ensuring that the enclave execution is {\em data oblivious}; i.e., the access pattern no longer depends on the input data~\cite{STC}. While \codename\ does not explicitly eliminate side-channel leakage, it could benefit from a vast amount of research on defenses against side-channel leakages~\cite{path_oram, oblivm, STC, page_fault_sgx, pdedup}, which we shall incorporate into \codename\ in future work.

\section{Evaluation}
\label{sec:eval}
This section reports empirical evaluation of our prototype implementation of \codename. We are interested in quantifying the overhead of enclave over non-enclave (and thus untrusted) execution (Section~\ref{subsec:enclave_cost}) and the cost of task matching (Section~\ref{subsec:task_matching_cost}).

\subsection{Experimental Setup}
All experiments are conducted on a system that is equipped with Intel i7-6820HQ 2.70GHz CPU, 16GB RAM, 2TB hard drive, and running Ubuntu 16.04 Xenial Xerus. We evaluate the overhead of \codename's enclave execution using a number of computational tasks including five benchmarks (i.e., mcf, deepsjeng, leela, exchang2, and xz) selected from SPEC CPU2017~\cite{spec_cpu}, and two standard cryptographic operations (i.e., SHA256 and AES Encryption). The enclave trusted codebases are implemented using Intel SGX SDK~\cite{sgx_sdk}. To quantify the cost of task matching in \codename, we measure the runtime of the Mucha-Sankowski algorithm~\cite{max_matching} that we implemented in C. All experiments are repeated over 10 runs, and the average results are reported.

\subsection{Cost of Enclave Execution}
\label{subsec:enclave_cost}
\noindent\textbf{Overhead in Execution Time.}
We evaluate the five SPEC CPU2017 benchmarks in three different execution modes, namely {\em baseline}, {\em SGX-compliant} and {\em \codename-compliant}. The baseline mode compiles the benchmarks as-is and runs them in untrusted execution environment. SGX-compliant mode requires porting the benchmarks to support SGX-enclave execution. This entails replacing standard system calls and libraries in the original code with SGX-compliant ones supported in the SGX SDK~\cite{sgx_sdk}. Finally, the \codename-compliant mode further instruments SGX-compliant code with dynamic runtime checks and TEE-based metering discussed in previous section. 

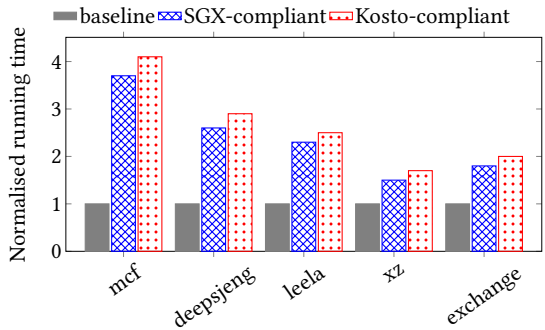
\begin{figure}[t]
\centering
\begin{tikzpicture}[thick, scale = 0.5]
\begin{axis}[
	width = 0.8\textwidth,
	height=\axisdefaultheight,
	ticklabel style = {font=\Huge},
    ybar,
    enlarge x limits=0.16,
    legend style={at={(-0.035,1.1)},anchor=west, draw=none,legend columns=3, font = \Huge},
    bar width=18pt,  
    area legend,
    ylabel={Normalised running time},   
    y label style={font = \Huge, at={(0.0,0.5)}},   
    xticklabels={mcf, deepsjeng, leela, xz, exchange},
    xtick=data,
    ymin = 0,
    xticklabel style={rotate=35},
]
\addplot [color = gray, thick, fill = gray] coordinates {
(1,1) 
(2,1) 
(3,1) 
(4,1)
(5,1)
};

\addplot  [color = blue,thick, pattern color = blue,pattern=crosshatch] coordinates {
(1,3.7) 
(2,2.6) 
(3,2.3) 
(4,1.5)
(5,1.8)
};

\addplot  [color = red,thick, pattern color = red,pattern=dots] coordinates {
(1,4.1) 
(2,2.9) 
(3,2.5) 
(4,1.7)
(5,2.0)
};

\legend{baseline, SGX-compliant, \codename-compliant}
\end{axis}
\end{tikzpicture}
\vspace{-1.5mm}
\caption{\codename's enclave execution overhead. The running time of each benchmark is normalized against its own baseline mode's.}
\vspace{-1.5mm}
\label{fig:cpu_spect_overhead}
\end{figure}

Figure~\ref{fig:cpu_spect_overhead} compares the running time of the five benchmarks in three modes, with the running time of each benchmark normalized against its own baseline. We observe that the SGX-compliant mode incurs from $1.5\times$ to $3.7\times$ overhead over the baseline. This overhead is mostly due to enclave's control switching. The instrumentations introduced in \codename-compliant mode incur an extra $8\% \sim 14\%$ overhead relative to the SGX-compliant mode. 

We remark that our porting of the five benchmarks to SGX might not be optimized. Thus, the results reported in Figure~\ref{fig:cpu_spect_overhead} are likely an \textit{over-estimation} of the real overhead that enclave execution incurs over unstrusted non-enclave execution. Various techniques have been proposed for minimizing the overhead of enclave execution, typically by reducing the control switching between the enclave code and the untrusted application that services OS-provided functions~\cite{eleos, vault, hotcalls}. We leave the incorporation of such optimization into \codename\ for future work.

\definecolor{bistre}{rgb}{0.24, 0.17, 0.12}
\definecolor{darkorange}{rgb}{1.0, 0.55, 0.0}
\definecolor{darkblue}{rgb}{0.0, 0.0, 0.55}
\begin{figure}[t]
\centering
\begin{subfigure}{.235\textwidth}
\begin{tikzpicture}[thick, scale = 0.48]
\begin{axis}[
	ticklabel style = {font=\Huge},
    legend style={at={(0.5,0.25)},anchor=west, draw=none,legend columns=1, font = \huge},
    ylabel={MBps},   
    y label style={font = \huge, at={(0.0,0.5)}},   
    xlabel={Message Size (Bytes)},  
	x label style={font = \huge}, 
    xtick=data,
    x label style={font = \huge, at={(0.5,-0.05)}},
    xticklabels={$2^{8}$, , $2^{12}$, , $2^{16}$, , $2^{20}$, ,$2^{24}$},
    ymin = 0,
]
\addplot [mark = triangle*, smooth,thick,  color = darkblue,mark options={scale=2, fill = darkblue}] coordinates {
(1,270) 
(2,348) 
(3,392) 
(4,408)
(5,432)
(6,441)
(7,442)
(8,441)
(9,443)
};

\addplot  [mark = *, smooth, thick, color = darkorange, mark options={scale=1.5, fill = darkorange}] coordinates {
(1,56) 
(2,135) 
(3,219) 
(4,253)
(5,298)
(6,301)
(7,303)
(8,303)
(9,302)
};

\legend{OpenSSL, SGXSSL}
\end{axis}
\end{tikzpicture}
\caption{SHA256 throughput}
\label{subfig:sha_throughput}
\end{subfigure}
\hfill
\begin{subfigure}{.235\textwidth}
\begin{tikzpicture}[thick, scale = 0.48]
\begin{axis}[
	ticklabel style = {font=\huge},
    legend style={at={(0.5,0.25)},anchor=west, draw=none,legend columns=1, font = \huge},
    ylabel={GBps},   
    xlabel={Message Size (Bytes)},   
    y label style={font = \huge, at={(0.05,0.5)}},   
    x label style={font = \huge, at={(0.5,-0.05)}},
    xtick=data,
	xticklabels={$2^{8}$, , $2^{12}$, , $2^{16}$, , $2^{20}$, ,$2^{24}$},
    ymin = 0,
]
\addplot [mark = triangle*, smooth,thick,  color = darkblue,mark options={scale=2, fill = darkblue}] coordinates {
(1,0.43) 
(2,1.38) 
(3,3.07) 
(4,4.22)
(5,4.89)
(6,5.02)
(7,5.01)
(8,5.06)
(9,5.04)
};

\addplot  [mark = *, smooth, thick, color = darkorange, mark options={scale=1.5, fill = darkorange}] coordinates {
(1,0.13) 
(2,0.22) 
(3,0.96) 
(4,2.13)
(5,3.78)
(6,4.66)
(7,4.87)
(8,4.92)
(9,4.91)
};
\legend{OpenSSL, SGXSSL}
\end{axis}
\end{tikzpicture}
\caption{AES-GCM throughput}
\label{subfig:aes_throughput}
\end{subfigure}
\vspace{-1.5mm}
\caption{Comparison between throughput of enclave and non-enclave based cryptographic operations.}
\vspace{-1.em}
\label{fig:throughput_overhead}
\end{figure}
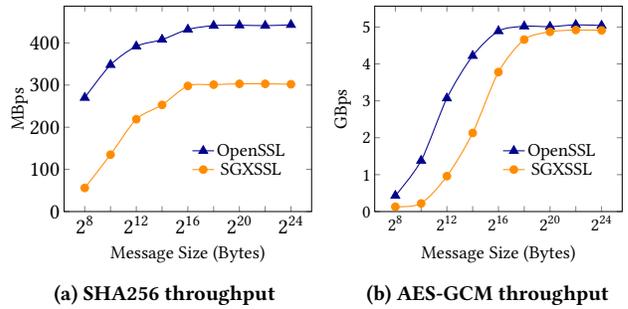

\noindent\textbf{Overhead in Throughput.} Next, we measure the overhead in throughput incurred by enclave execution on computation-intensive works. This set of experiments measure performances of  SHA256 and AES-GCM encryption operations under \textit{OpenSSL}~\cite{openssl} and Intel \textit{SGXSSL}~\cite{sgxssl} implementations against exponentially increasing input size (ranging from 256B to 4MB). 
OpenSSL implementation runs in an untrusted non-enclave memory, whereas SGXSSL ports OpenSSL to support SGX enclave execution. 

Figure~\ref{subfig:sha_throughput} shows a significant gap between the throughput of SGXSSL and OpenSSL implementations of SHA256  when a message size is small (e.g., OpenSSL's throughput is upto $5\times$ for 1KB message). Nonetheless, such a gap reduces as the message size increases (e.g., as small as $1.5\times$ for 4MB message). A similar trend is observed in throughput of AES-GCM encryption (the decryption throughput is similar), with the throughput overhead incurred by enclave execution reduces from $6.3 \times$ for 1KB message to $3\%$ for 4MB message. We attribute this throughput gap to the I/O cost and context switching that enclave execution incurs. Fortunately, this overhead is amortized as the input size increases.

\subsection{Cost of Task Matching}
\label{subsec:task_matching_cost}
Finally, we evaluate the performance of \codename\ task assignment by measuring the runtime of Mucha-Sankowski algorithm. The algorithm runs in $O(|V|^{2.38})$, where $|V|$ is the total number of vertices in the graph.

Table~\ref{tab:mbp} reports running time of the Mucha-Sankowski algorithm on dense bipartite graphs whose densities  (i.e., $D = \frac{2|V|}{|V|(|V|-1)}$) range from $0.8$ to $0.9$. As expected, the running time grows quadratically with respect to the input size. 
This limits the  task assignment algorithm to offer real-time performance only when the number of requests and compute nodes are small (e.g., 4000). To offer real-time performance at scale, \codename\ expects a number of brokers offer competing services. Alternatively, a single broker can trade the optimality of the task assignment for lower cost of task matching by splitting requests and compute nodes into multiple batches, and evaluating the batches in parallel.

\begin{table}[]
\centering
\caption{Running time of Mucha-Sankowski algorithm.}
\vspace{-2mm}
\label{tab:mbp}
\begin{tabular}{c|ccccc}
\hline
\textbf{|V|} & 1000  & 2000 & 4000  & 8000 & 16000    \\
\hline
\textbf{Running time (s)} & 0.07  & 0.43    & 1.83   & 9.72   & 47.96 \\
\hline
\end{tabular}
\vspace{-1em}
\end{table}

\section{Related Works}
\label{sec:related_works}
\textbf{Decentralised outsourced computation}. Golem~\cite{golem} also explores a marketplace for outsourced computation. Unlike \codename, it does not feature the attested execution environment. Consequently, Golem needs to redundantly execute the same task on multiple compute nodes in order to verify the execution correctness.

Concurrent to our work, AirTNT~\cite{airtnt} proposes the use of enclave execution for outsourced computations, and devises a protocol that allows fair exchange between the client and the compute nodes. Such protocol necessitates a separate payment channel for every pair of client and compute node, and requires constant communication between the two parties over the course of the outsourced computation (i.e., highly interactive). \codename, in contrast, alleviates the client and the compute nodes from these inconveniences. 

\textbf{CryptoCurrency and Off-Chain Payment Channel}. Cryptocurrencies allow two willing parties to transact directly via an open and decentralized blockchain~\cite{btc_origin}. 
Beyond a means of transacting, blockchain architectures have been developed for various purposes, such as enabling a Turing-complete smart contract platform~\cite{eth_origin} or enhancing transaction~\cite{zcash, monero} privacy. 
We refer readers to \cite{cryptocurrency_sok} for a comprehensive overview of cryptocurrencies.

Current prominent cryptocurrencies can only support limited transaction throughputs. As a result, various off-chain payment solutions have been presented. Lighting network~\cite{lightning_bitcoin} and Raiden network~\cite{raiden_eth} are earlier off-chain payment solutions proposed for Bitcoin and Ethereum, respectively. Sprites~\cite{sprites} and Revive~\cite{revive} optimize the costs of indirect off-chain payments, while Bolt~\cite{bolt} allows privacy-preserving off-chain payment channels. \codename\ design has thus far only tapped on a basic feature of payment channels, and can certainly benefit from their further developments.

\textbf{Reliable Resource Accounting}. Early approaches to resource accounting in the context of outsourced computations rely on nested virtualization and TPMs, or place a trusted resource observer underneath the service provider's software~\cite{sekar2011verifiable, chen2013towards}. Alternatively, REM~\cite{REM} instruments the client's program with dynamic runtime checks that maintain an instruction counter to self account its computational effort. The correctness and integrity of these runtime checks are enforced by the trusted hardware. \codename\ adopts REM's approach in metering the compute nodes' work.

\textbf{SGX-based systems}. Trusted hardware, in particular Intel SGX processors, have been used to enhance security in various application domains, including data analytics~\cite{schuster2015vc3, m2r, STC}, machine learning~\cite{ohrimenko2016oblivious} and outsourced storage~\cite{podr, pdedup}. In addition, SGX has also been utilized to scale the blockchain~\cite{trustchain, poet}. To our knowledge, \codename\ is the first solution to provision a full-fledged fair marketplace for secure outsourced computations using Intel SGX.

\section{Conclusion}
\label{sec:conclusion}
We have presented \codename\ -- a framework enabling fair marketplace for secure outsourced computations. \codename\ protects confidentiality of clients' input, integrity of the computations, and ensures fair exchange between the clients and the compute nodes. 
Our experiments show that \codename\ incurs small overheads, as low as $3\%$ for computation-intensive operations, and $1.5\times$ for I/O-intensive operations over non-enclave and untrustworthy execution. 

\bibliography{paper}

\end{document}